\documentstyle{camera}

\newcommand{\AmS}{{\protect\the\textfont2
  A\kern-.1667em\lower.5ex\hbox{M}\kern-.125emS}}

\newcommand{\beq}{\begin{equation}}
\newcommand{\eeq}{\end{equation}}
\newcommand{\bea}{\begin{eqnarray}}
\newcommand{\eea}{\end{eqnarray}}

\def\dm2{\Delta m^2}
\def\sq2{sin^2(2\Theta)}

\input epsf

\begin{document}

%
\title{Structures, Oscillations and Solitons \\ in 
  Large-Scale Multi-component Self-Gravitating Systems}

%
\author{KINWAH WU$^1$, ZIRI YOUNSI$^{1,2}$ \And CURTIS J. SAXTON$^1$}

%
\organization{
1. Mullard Space Science Laboratory, University College London, 
 Holmbury St Mary, Surrey RH5 6NT, United Kingdom \\ 
 2. Department of Physics and Astronomy, University College London, 
  Gower Street, London WC1E 6BT, United Kingdom}

\maketitle

\begin{abstract}
We investigate the structure of dynamics of large self-gravitating astrophysical systems 
  using a self-interacting two-component model. 
We consider two cases, galaxy clusters and cosmic walls, for illustrations. 
In both cases stability analyses are conducted using perturbative expansion.  
We have found that waves and solitons are easily generated in these systems. 
Our analysis shows that dark matter can be Jeans unstable 
   in the very inner regions of galaxy clusters  
   if it has a large internal degree of freedom.     
The dark matter core may collapse under external perturbations.  
We also discuss dark-matter oscillations in galaxy clusters 
   and how mode growth and decay lead to heating of intracluster medium.   
Our analysis shows that dark-matter solitons with both positive and negative amplitudes 
   can be excited in cosmic walls. 
Resonances in soliton interaction could enhance gas condensation. 
The co-existence of the two types of dark-matter solitons 
   implies that bright filaments can arise in dark voids. 

\end{abstract}
\vspace{1.0cm}

\section{Multi-component Self-gravitating Systems} 

The Universe is a hierarchy of self-gravitating multi-component systems: 
  from walls, super clusters, galaxy clusters and groups    
   to galaxies, star clusters, stars and sub-stellar objects.  
The dark energy provides a repulsive pressure on very large scales,  
  but dark matter and baryons adhere gravitationally to form smaller structures.  
Single-component systems and multi-component systems possess very different properties 
  because of the complexities in dynamical interactions between the components, and 
  between the components and the environment.    
For instance, dark matter does not radiate electro-magnetic waves  
   but participates in gravitational interaction with baryons (gas).  
However, the baryonic gas, which has a much smaller inertia than the dark matter,  
   regulates the energy exchange of the system with the environment.  
We are still unsure as to the nature of the dark matter 
  and how many dark-matter species are present.  
We are also not certain if dark matter is strongly self-interacting or essentially collisionless.  
   
Here, we report the recent finding of our study of two example 
   multi-component self-gravitating systems: galaxy clusters and cosmic walls.      
We show the observational characteristics of galaxy clusters in our two-compoent model 
   with self-interacting dark-matter. 
We also demonstrate the formation of solitary waves in cosmic walls 
   when dark matter is self-interacting.     

\section{Galaxy Clusters} 

\subsection{Self-interacting Multi-component Model}

Galaxy clusters consist of about 85\% dark matter and 10\% hot ionised gas in mass 
  (see e.g.\ Vikhlinin et al. 2006). 
The trapped baryons in stars and galaxies have an insignificant mass contribution, 
  and thus play a less important role in the cluster dynamics. 
For simplicity, If we ignore the magnetic fields, cosmic rays and galactic outflow, 
  galaxy clusters can be treated as two-component systems 
  consisting only of dark matter and baryons.  
The cluster structure and dynamics is then determined by the conservation equations:  
\begin{eqnarray} 
  \frac{\partial}{\partial t} \rho_{\rm i} + \nabla \cdot \rho_{\rm i} {\mathbf v}_{\rm i}   
    & = &  0  \ ;  \\  
  \frac{\partial}{\partial t} \rho_{\rm i} {\mathbf v}_{\rm i} 
    + \nabla \cdot \rho_{\rm i} {\mathbf v}_{\rm i} {\mathbf v}_{\rm i} 
     + \nabla \rho_{\rm i} \sigma_{\rm i}^2
     & = &  \rho_{\rm i} {\mathbf f} \ ;  \\ 
   \frac{\partial}{\partial t}s_{\rm i} + {\mathbf v}_{\rm i} \cdot \nabla s_{\rm i} 
     & = &  (\gamma_{\rm i}- 1)\rho_{\rm i}^{-\gamma_{\rm i}} \cal{L}_{\rm i}   
\end{eqnarray}  
  (i = 1, 2 for gas and dark matter respectively), 
  where ${\mathbf v}_{\rm i}$ the bulk velocity and ${\cal L}_{\rm i}$ the energy gain/loss rate.      
For self-interacting dark matter and gas, the effective pressure $p_{\rm i}$, 
   density $\rho_{\rm i}$, velocity dispersion  $\sigma^2_{\rm i}$ 
   and specific entropy $s_{\rm i}$ are related by   
   $p_{\rm i} = \rho_{\rm i} \sigma_{\rm i}^2 = s_{\rm i} \rho_{\rm i}^{\gamma_{\rm i}}$, 
   where $\gamma_{\rm i}$ is the adiabatic index and 
   the degree of freedom ${F_{\rm i}} = 2/ (\gamma_{\rm i} - 1)$.    
The gravitational force $\mathbf{f} = - \nabla \phi$, 
   where the potential $\phi$ satisfies the Poisson Equation.  

\begin{figure}[h!]  
\epsfxsize=4.8cm \hspace{0.65cm} \epsfbox{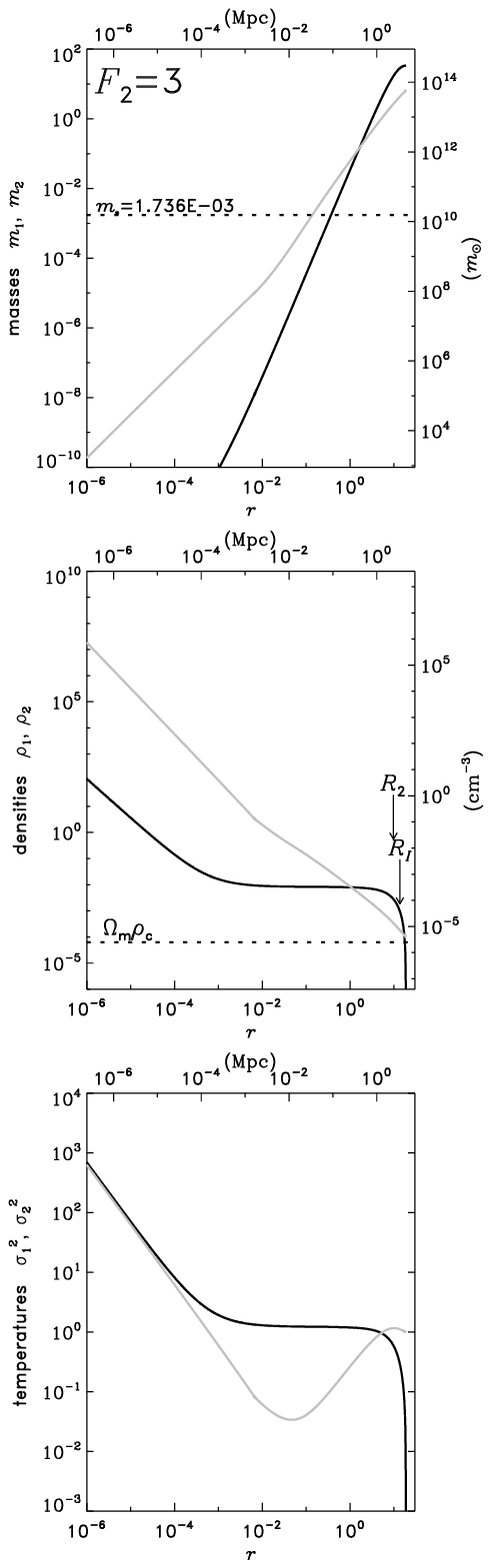} 
\epsfxsize=4.8cm \hspace{0.7cm} \epsfbox{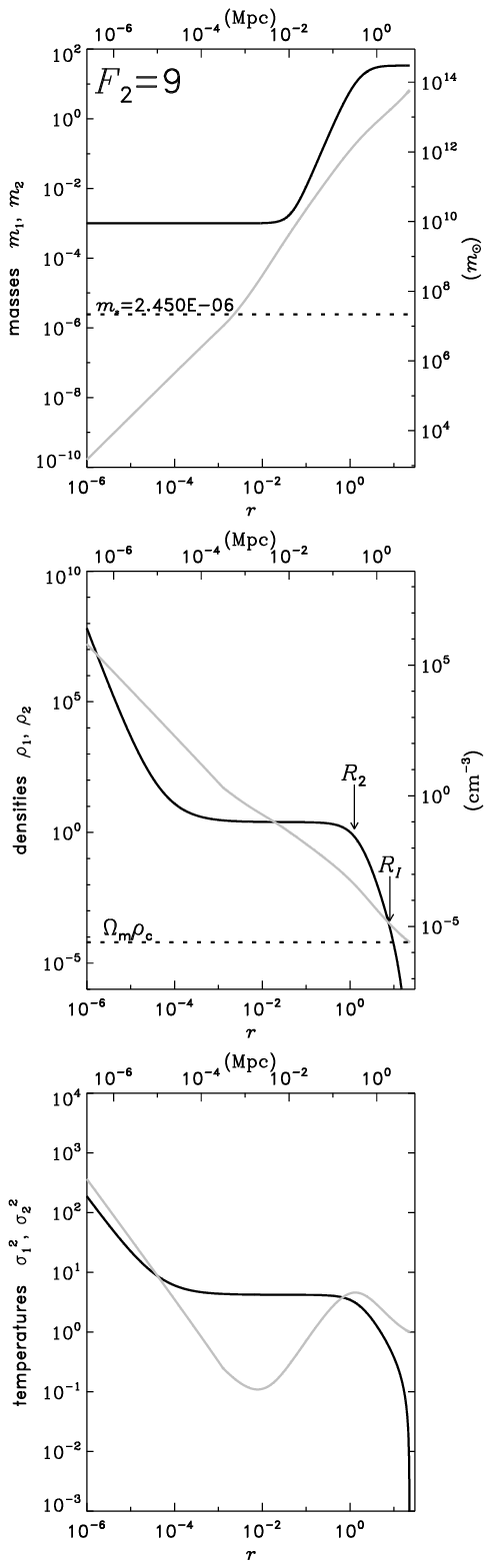} 
\vspace{0.2cm} 
\caption[h]{Stationary mass density and temperature profiles of model clusters 
  with$F_2 = 3$ and 9 (left and right columns respectively). 
  The mass inflow rate $\dot M = 10~ M_\odot{\rm yr}^{-1}$ and 
  the temperature at the outer boundary $T_R =1$~keV. 
Solid black lines correspond to dark-matter profiles and grey lines, gas profiles. 
The dotted lines indicate the cosmic mean density.   }
\end{figure}
   
For stationary clusters, $\partial/\partial t \rightarrow 0$, 
  and the radial structure profiles of the cluster is determined by    
\begin{equation} 
 \left[  
 \begin{array}{ccc} 
  v_{\rm i} &   \rho_{\rm i}  & 0 \cr 
  \sigma_{\rm i}^2   & \rho_{\rm i} v_{\rm i}  &  \rho_{\rm i }\cr   
  0  &  \rho_{\rm i} v_{\rm i}^2 & \frac{\gamma_{\rm i}}{\gamma_{\rm i}-1}
   \rho_{\rm i} v_{\rm i}  
 \end{array}   
 \right] 
 \frac{d}{dr} 
 \left[ 
 \begin{array}{c} 
  \rho_{\rm i} \cr 
  v_{\rm i} \cr 
  \sigma_{\rm i}^2  
 \end{array}  
\right] 
 = 
\left[
\begin{array}{c}  
-\frac{2}{r}\rho_{\rm i} v_{\rm i} \cr 
 \rho_{\rm i} f  \cr 
 \rho_{\rm i} v_{\rm i} f + {\cal L}_{\rm i}
\end{array} 
\right] \  . 
\end{equation}    
The Poisson Equation gives 
\begin{equation} 
 \frac{1}{4\pi G} \left[ \frac{d}{dr} f  + \frac{2f}{r} \right] = \sum_{\rm i} \rho_{\rm i}   \ , 
\end{equation} 
  where $G$ is the gravitational constant 
  (see Saxton and Wu 2008). 
No radiative loss from dark matter implies ${\cal L}_2 = 0$. 
For the radiative loss from gas, a parametrisation    
  ${\cal L}_1 = A \rho_1^\alpha (\sigma_1^2)^\beta$ is appropriate.   
Assume that free-free emission is the dominant cooling process.  
Then we have $\alpha =2$ and $\beta = 1/2$, 
  and $A$ is set by the chemical abundances. 
The presence of radiative cooling of gas causes a pressure depletion in the cluster interior, 
  leading to a gas inflow, at a rate $\dot M = 4\pi r^2 \rho_{\rm i} v_{\rm i}$.

\begin{figure}[h!]  
\epsfxsize=4.8cm \hspace{0.65cm} \epsfbox{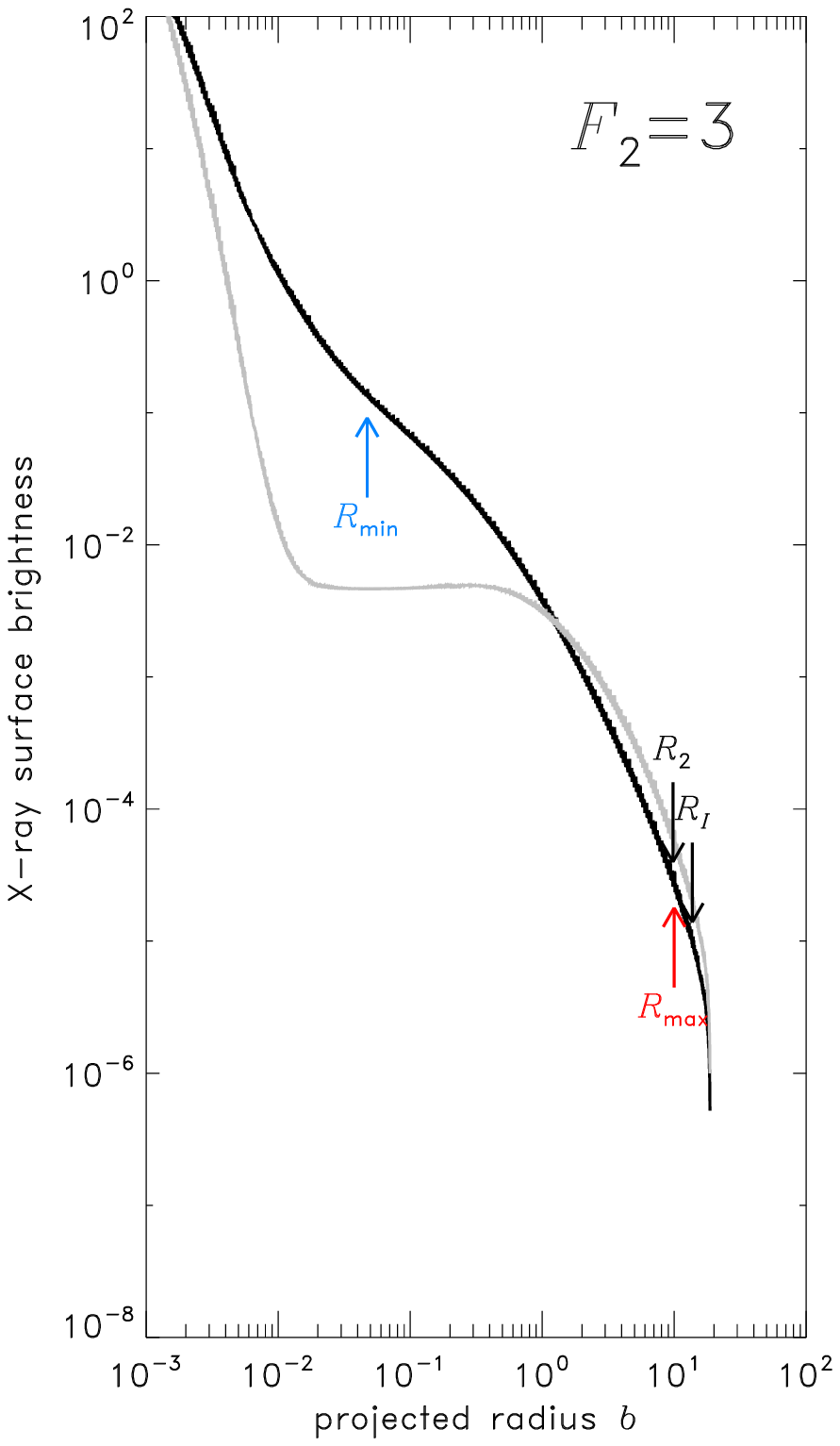} 
\epsfxsize=4.8cm \hspace{0.7cm} \epsfbox{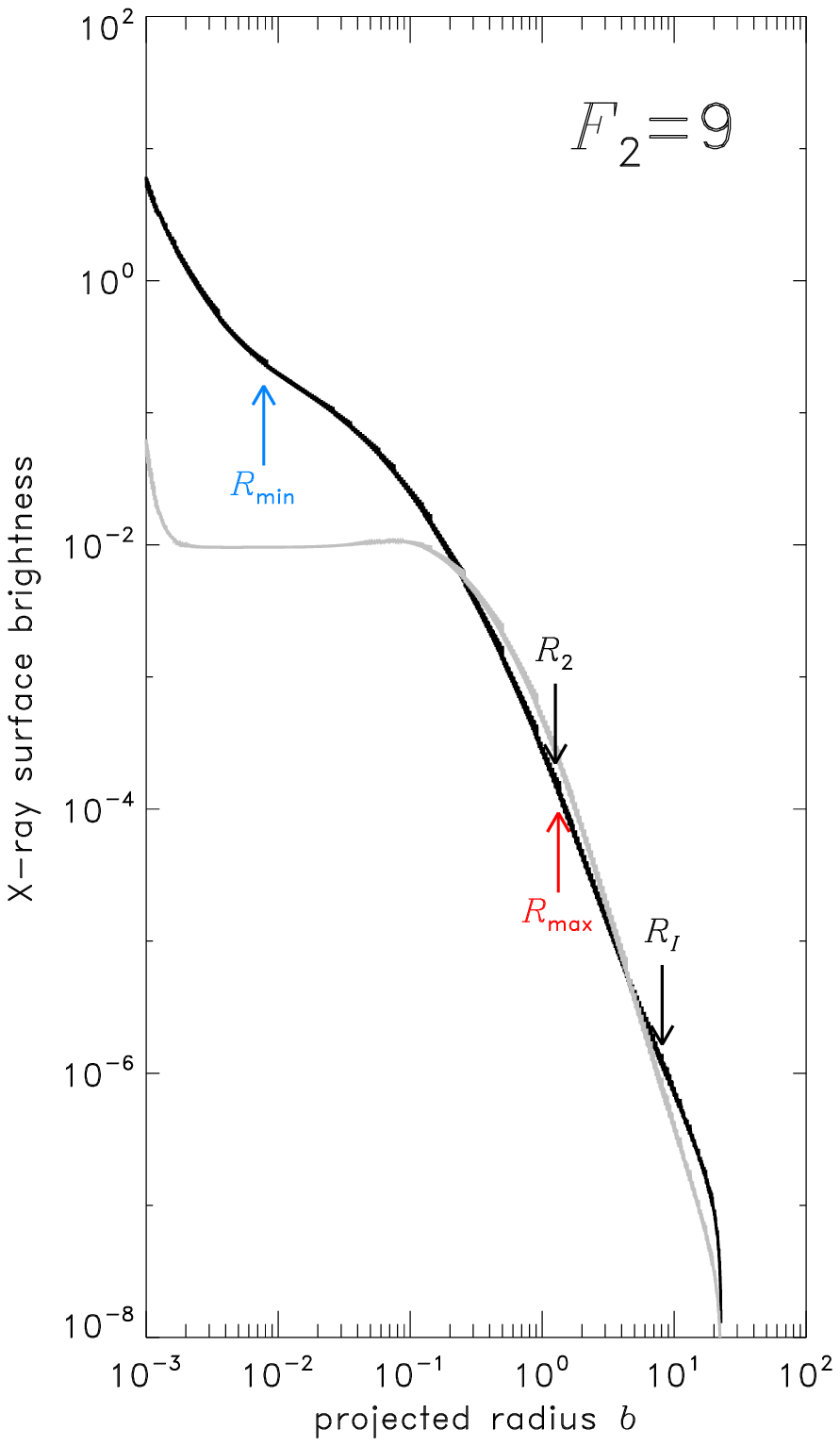} 
\vspace{0.2cm} 
\caption[h]{X-ray surface brightness profiles of the two clusters in Fig.~1. 
   Black lines indicate emission at the $0.1-2.4$~keV band, 
      and gray lines, the $2-10$~keV band. }
\end{figure}

\begin{figure}[h!]  
\epsfxsize=4.8cm \hspace{0.65cm} \epsfbox{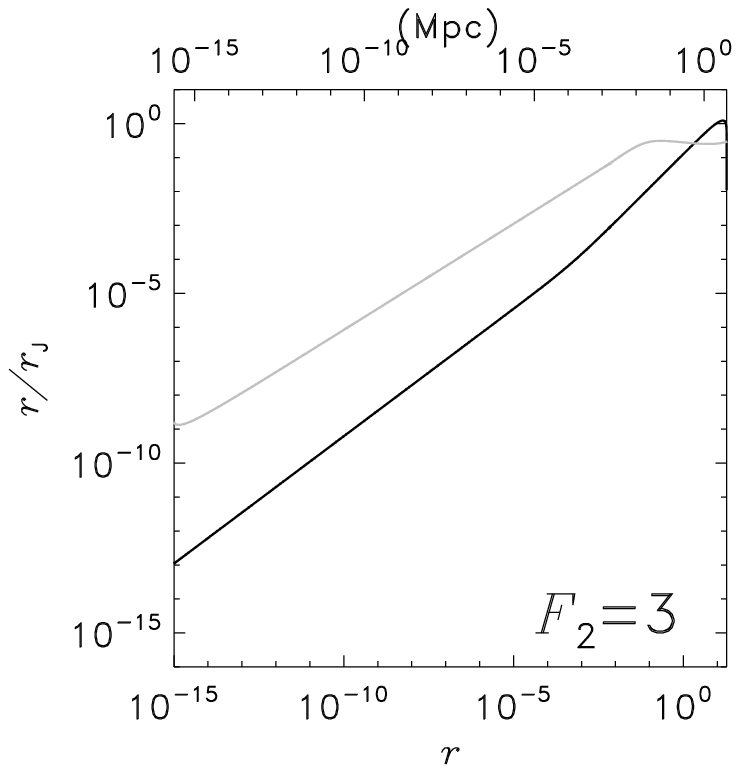} 
\epsfxsize=4.8cm \hspace{0.7cm} \epsfbox{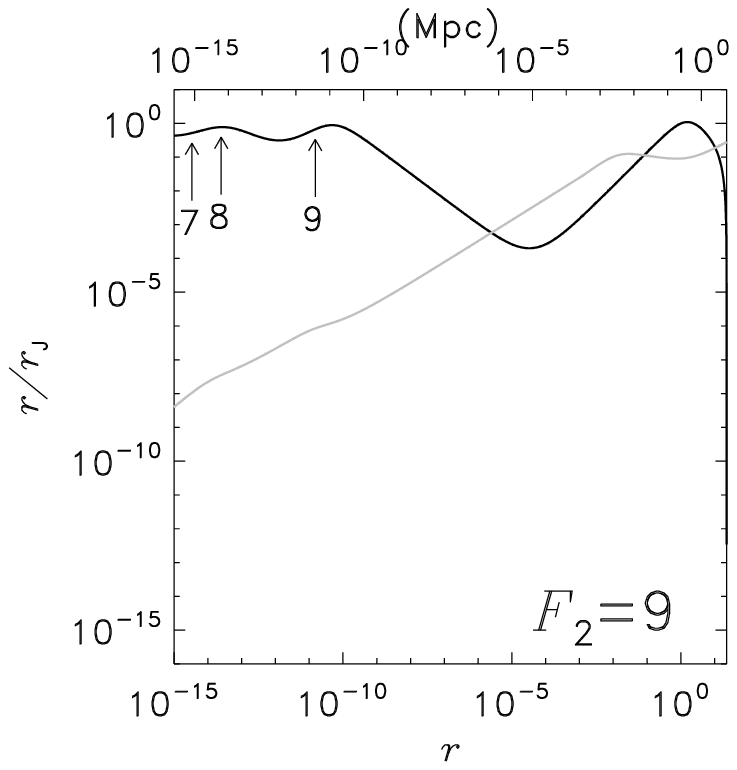} 
\vspace{0.2cm} 
\caption[h]{The ratio of the radius to the local Jean length for the clusters in Fig.~1. 
  Black lines indicate dark matter, and gray lines, gas. 
  Arrows with labels 7, 8 and 9 mark the marginally stable halos 
     enclosing dark-matter mass of $10^7$, $10^8$ and $10^9$~M$_\odot$.   }
\end{figure}

\subsection{Stationary Structures, Stability and Observational Implications}

With the boundary conditions specified, Equations (4) and (5) can be integrated. 
Two example structures of clusters with different $F_2$ are shown in Figure 1.    
For $F_2 = 3$, the gas inertia is important in the cluster core; 
  for $F_2 = 9$, dark matter dominates throughout the cluster.  
In both cases, the temperature has a depression at about $10^{-2}$ of the cluster radius 
  and increases at larger radii.  
For $F_2 = 9$, the temperature first reaches a peak 
  and then has a down-turn at the cluster out-skirt. 
The temperature down-turn at the cluster out-skirt is less obvious 
   in the cluster with $F_2 = 3$.  
Figure 2 shows the X-ray surface brightnesses of the clusters.  
A bright hard X-ray core is present for $F_2 =3$, but is absent for $F_2 = 9$.    
Figure 3 shows the local Jean lengths.  
For $F_2 =2$ both dark matter and gas are stable under the Jean's criterion.   
When $F_2 = 9$, gas is stable throughout the cluster 
  but dark matter can be unstable at the cluster core region.  
This implies that gas-free dark-matter collapse may occur 
  under certain structure perturbations (Saxton and Wu 2008).  

To investigate the global stability properties of the cluster, 
  we consider perturbations to the stationary structure and carry out a linear analysis.  
The Lagrangian perturbation is used,  
  in which the hydrodynamic variables take the form: 
\begin{eqnarray} 
  X_{\rm k} (\xi) & \rightarrow &  X_{\rm 0k}(\xi) 
    \left[1 + \epsilon \lambda_{\rm k} (\xi) e^{i(\kappa \xi - \omega - \varphi_{\rm k})} \right]   \  .  
\end{eqnarray} 
With this perturbative expansion, we obtain a set of differential equations, 
  which are solved by direct integration 
  after appropriate boundary conditions are imposed.  
The dimensionless eigenvalue of the dynamic equations is given by
\begin{eqnarray}  
   \delta & = & i \omega ~\left( \frac{X_0(\xi)}{\dot{X_0 (\xi)}} \right)   \  .  
\end{eqnarray}   
The eigen-plane plots show sequential modes. 
The imaginary part of the eigenvalue  ($\delta_{\rm I}$)   
  alternates between positive and negative values. 
These positive and negative $\delta_{\rm I}$ 
  correspond to the growth and decay modes of the disturbances in the cluster.  
Our calculations show a trend that the strengths of the modes increase with the mode orders, 
  implying that the higher-order modes 
  grow or decay faster than the lower-order modes.  

The presence of radiative cooling of gas in galaxy clusters are expected to cause cooling gas inflow 
  (Cowie and Binney 1977; Fabian, Nulsen and Canizares 1984). 
However, this phenomenon has not been confirmed by observations.   
It has been speculated that the gas inflow may be halted 
 as the radiative cooling may be compensated by certain efficient heating mechanisms.  
One argument  is that 
  hot electrons are injected into the intracluster medium by some means, 
  e.g.\ jets from AGNs residing in the cluster (McNamara et al.\ 2005). 
However, this would require some fine tuning. 
If the AGN injection is too strong the intra-cluster gas might be blown away; 
  if the AGN injection is too weak it might be sufficient to quench the cooling flow.   
An alternative is the excitation of a cluster tsunami which could heat up the cool cluster core 
  (Fujita, Suzuki and Wada 2004; Fujita et al.\ 2005). 
In the original tsunami model, the baryonic gas is trapped  
  in a potential well created by a static dark matter halo. 
The sound speed of gas is high in hot outer cluster region  
  than is much lower in the cool cluster core. 
Disturbance (waves) generated in the outer cluster region may be subsonic initially, 
  but when it propagates into the cooler cluster core, they could become supersonic. 
The waves will therefore pile up at the cool core region, leading to wave tsunami and shocks. 
The dissipation in the shocks presumably causes heating and quench the cooling flow.   

Our analysis provides a more self-consistent cluster tsunami scenario, 
  in that the dark matter halo provides a gravitational potential to trap the gas 
  as well as acts a oscillator that drives the waves in the gas. 
The piling up of waves would occur as in the original tsunami model, 
  but the tsunami is forced by the dark matter oscillations instead of self-excited. 
Our finding that higher-order modes grow faster   
  implies that long wave-length oscillations will break down into shorter wave-length oscillations.  
Thus, energy in the global cluster low frequency oscillation 
  will be transfered to smaller scale oscillations, causing tsunamis and shocks 
  which dissipate and heat the cluster core efficiently and evenly.   

\begin{figure}[t!]  
\begin{center} 
\epsfxsize=8.5cm \hspace*{0cm} \epsfbox{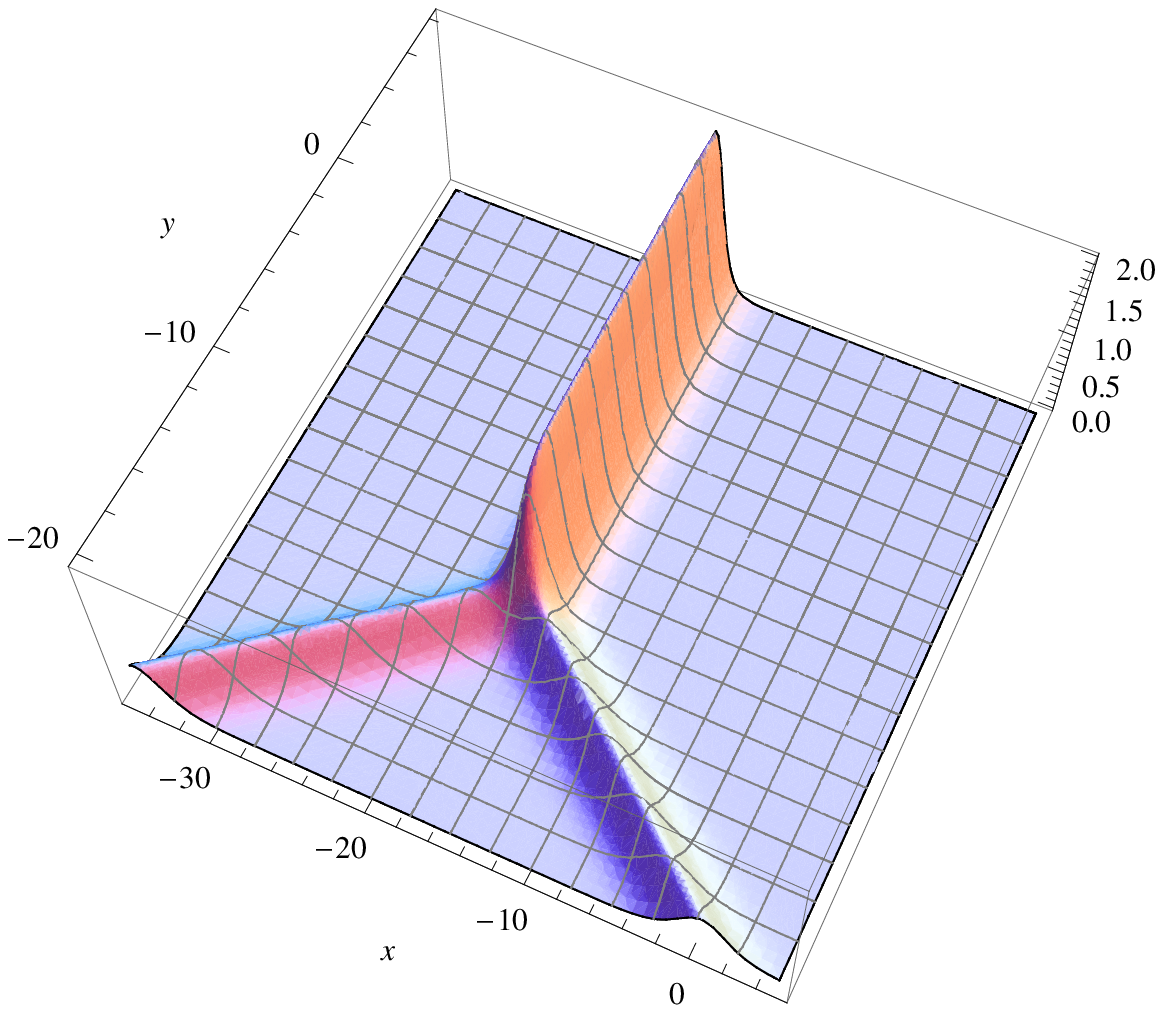}  
\end{center}
\vspace{0.cm} 
\caption[h]{Collision of two dark-matter solitons in the quasi-1D two-component cosmic wall. 
   The two dark-matter solitons have equal amplitude (height) initially. 
   They collide to form a resonant soliton 
      which has an amplitude four times that of the incident solitons. 
   This resonant dark-matter soliton has an infinite life-time. 
    As the dynamical time of the baryonic gas, i.e.\ $\tau \sim ~(4\pi G \rho_1)^{-1/2}$, is finite,  
     enhanced gas trapped and condensation will result from the formation of such resonant solitons.   }
\end{figure}

\section{Cosmic Walls}

\subsection{Hydrodynamic Model}   

Cosmic walls are semi-infinite two-component self-gravitating sheets.  
For self-interacting dark matter and baryonic gas, the equation of the state is   
$p_{\rm i} = s_{\rm i} \rho_{\rm i}^{\gamma_{\rm i}}=\rho_{\rm i}\sigma_{\rm i}^2$ 
   (here i = 1 for gas and i = 2 for dark matter).  
Since the dynamics are dominated by the dark matter, 
   we may ignore the inertia of the baryonic component 
   and simplify the conservation equations:    
\begin{eqnarray} 
  \frac{1}{\rho_1}\nabla p_1 
        & = & -  \nabla \phi  \ ; \\ 
  \frac{\partial}{\partial t} \rho_2 + \nabla \cdot \rho_2 {\mathbf v}_2   
    & = &  0  \ ;  \\  
  \frac{\partial}{\partial t} {\mathbf v}_2 
     + ( {\mathbf v}_2 \cdot \nabla   ) {\mathbf v}_2 - \frac{1}{\rho_2}\nabla p_2 
        & = &  -\nabla \phi \ .  
\end{eqnarray}  
The Poisson equation is given by 
\begin{eqnarray} 
 \nabla^2 \phi & = & 4\pi G \left(  \sum_{\rm i} \rho_{\rm i} - {\rho}_0 \right) \ ,   
\end{eqnarray}   
 where  $\rho_{\rm i}$  are the deviations 
 from the equilibrium background density ${\rho}_0$.  

\subsection{Soliton Formation} 

Cosmic walls are sheet-like, and we use a quasi-1D treatment, such that    
  $\partial/\partial x \gg \partial/\partial y > \partial/\partial z = 0$.  
We expand the density, velocity and potential 
   in terms of a perturbative parameter $\varepsilon$, i.e.\    
\begin{eqnarray} 
\rho & = & \rho^{(0)} + \sum_{\rm j} \varepsilon^{\rm j} \rho^{\rm (j)} \ ; \\ 
 v & = &  \sum_{\rm j} \varepsilon^{\rm j} v^{\rm (j)} \ ; \\ 
 \phi & = &  \sum_{\rm j} \varepsilon^{\rm j} \phi^{\rm (j)} \ ,    
\end{eqnarray} 
  and consider stretched co-ordinates   
  $\lambda  \equiv  \varepsilon^{1/2} (x- c_{\rm s}t)$, and  
  $\tau  \equiv   \varepsilon^{3/2} t $.      
Without loss of generality, we assume an isothermal gas component 
   and set $\gamma_1 = 1$.     
The zeroth order expansion gives the effective sound speed    
\begin{equation}
 c_{\rm s} = \sqrt{ \gamma_2 s_2 (\rho^{(0)})^{\gamma_2 -1} 
  +\xi \sigma_1^2  } \  , 
\end{equation} 
  where $\xi = \rho^{(0)}/(\rho_0 -\rho^{(0)})$. 
The first order non-linear expansion yields 
\begin{eqnarray} 
 \frac{1}{c_{\rm s}} \frac{\partial \rho^{(1)}}{\partial \tau} + 
  \Omega \left( \frac{\rho^{(1)}}{\rho^{(0)}}   \right)   \frac{\partial \rho^{(1)}}{\partial \lambda}  
   -  \left(\frac{\xi ^2\sigma_1^4}{8\pi G \rho^{(0)}c_{\rm s}^2} \right) 
    \frac{\partial^3 \rho^{(1)}}{\partial \lambda^3}  & = & 0 \  , 
\end{eqnarray} 
where 
\begin{eqnarray} 
   \Omega & = & \frac{1}{2 c_{\rm s}^2}
 \left[  \gamma_2(\gamma_2 +1)s_2 (\rho^{(0)})^{\gamma_2 -1} 
 + \xi (\xi +3) \sigma_1^2  \right] \ .  
\end{eqnarray} 
Rescaling the variables leads to the KdV (Korteweg de-Vries) Equation: 
\begin{eqnarray}  
  \frac{\partial \bar{\rho}^{(1)}}{\partial \tau'} +  
      {\bar \rho}^{(1)}   \frac{\partial {\bar \rho}^{(1)}}{\partial \lambda'}  
   -  \mu   \frac{\partial^3 {\bar \rho}^{(1)}}{\partial \lambda'^3}  & = & 0 \   , 
\end{eqnarray}    
   where the parameter $\mu$ is determined 
   by the thermodynamic properties of the dark matter and the gas.  
 
The quasi-1D approximation can be relaxed 
  by  introduction of an additional transverse co-ordinate $y$ 
  and its associated perturbation $\eta = \varepsilon y$.  
The perturbation of the according velocity component would take the form:      
\begin{eqnarray} 
   v_y & = & \sum_{\rm j} \varepsilon^{\rm j+\frac{1}{2}} v_y^{\rm (j)} \ .  
\end{eqnarray}   
With rescaling of the variables, 
  the first order non-linear expansion yields   
\begin{eqnarray}  
   \frac{\partial}{\partial \lambda'}
    \left[  \frac{\partial \bar{\rho}^{(1)}}{\partial \tau'} +  
   \Xi  {\bar \rho}^{(1)}   \frac{\partial {\bar \rho}^{(1)}}{\partial \lambda'}  
   -  \mu   \frac{\partial^3 {\bar \rho}^{(1)}}{\partial \lambda'^3} \right]  
     +\frac{\partial^2 {\bar \rho}^{(1)}}{\partial \eta'^2} & = & 0 \  , 
\end{eqnarray}  
    which is the KP (Kodomstev-Petviashvili) Equation.  
The parameters $\Xi$ and $\mu$ are determined by the thermodynamics 
  of the dark matter and the gas.  
  
Note that the isothermal gas assumption, i.e.\ $\gamma_1 = 1$, 
  is not essential for the derivation of the KdV and the KP Equations. 
The two equations can be obtained using a polytropic equation of state for the gas 
  (Younsi 2008).   

\subsection{Astrophysical Implications}  

A family of solutions to the KdV Equations and the KP Equations 
  are the solitary waves (see e.g. Belashov and Vladimirov 2005). 
Solitary waves are non-linear waves, and they interact non-linearly.  
Linear superposition 
  predicts an amplitude doubling when two linear waves  with equal amplitudes interact.   
However, collision of two solitons with equal amplitudes 
  could form a metastable resonant pulse 
  whose amplitude is four times that of the incident solitons.    
This non-linear superposition has important implications 
  in the dynamics and structure formation of large-scale astrophysical systems.    
For instance, when the resonant soliton resulting from the collisions of dark-matter solitions 
  of positive densities  causes a depression in the local gravitational potential, 
  if the resonant time-scale is larger than the dynamical time of the baryonic gas, 
  the gas will be trapped into this depression. 
Once the Jeans instability threshold is reached, 
  the trapped gas will collapse and condense into visible structures.  
As there is a larger amplitude enhancement in the soliton interactions, 
  it increases the efficiency of gas trapping and hence the subsequent condensation.  
  
In the quasi-1D case, we obtain a dispersion relation as   
\begin{eqnarray} 
 \omega & = & k c_{\rm s} \left( 1+  
   \frac{k^2(\xi \sigma_1^2)^2}{c_{\rm s}^2 [4\pi G \rho^{(0)} - k^2(\xi \sigma_1^2)] } 
      \right)^{1/2} \ .   
\end{eqnarray} 
It follows that the Jeans wave number 
\begin{eqnarray} 
  k^2_{\rm J} & = & \frac{4\pi G \rho^{(0)}}{c_{\rm s}^2 - \xi \sigma_1^2}  
     \left( \frac{c_{\rm s}^2}{\xi \sigma_1^2} \right) 
      = \frac{4\pi G(\rho_0 - \rho^{(0)})}{\gamma_2 s_2 (\rho^{(0)})^{\gamma_2 -1}} 
          \left(\frac{c_{\rm s}^2}{\sigma_1^2} \right) \ , 
\end{eqnarray}  
  which is always positive for $\gamma_1 >0$.  
From the dispersion relation, one can see 
   that the system is stable for  $0< k < (4\pi G \rho^{(0)}/\xi \sigma_1^2)^{1/2}$   
   and for $k> k_{\rm J}$, 
   marginally stable for $k=0$ and $k= k_{\rm J}$,  
   but unstable for $(4\pi G \rho^{(0)}/\xi \sigma_1^2)^{1/2} < k < k_{\rm J}$. 
As $\gamma_1 >$, 
  bright solitons are formed when $k> k_{\rm J}$ 
  but dark solitons are formed for $k < (4\pi G \rho^{(0)}/\xi \sigma_1^2)^{1/2}$
  (Younsi 2008).  
The bright solitons correspond to dark matter density pulses 
  with a positive amplitude above the equilibrium background density,  
  and the dark solitons correspond to dark matter density pulses 
  with a negative amplitude. 
The two types of solitons can co-exist locally in the presence of high-order perturbations, 
  leading to the formation of `double layers' 
  (Younsi 2008; see also Shukla and Mamun 2002).    
The formation of weak filaments (galaxies and galaxy clusters) deep inside 
  a cosmic void has been a challenging issue in astrophysics. 
Conventional wisdom predicts no growth of structure inside large scale low-density regions, 
  contrary to observation. 
The formation of `double layer' soliary waves in two-component self-graviating sheets 
  may provide a viable model for the formation of such filament and void structures.   

\section{Summary}

We construct a two-component model for large-scale self-gravitating systems. 
The two components are self-interacting.  
Two example cases, galaxy clusters and cosmic walls, are presented. 
Our key findings are summarised as follows. 
When the dark-matter has a large internal degree of freedom, 
  it can be Jeans unstable locally, implying the possibility of collapse under external perturbations. 
Dark matter oscillations in galaxy clusters can lead to gas tsunamis in cluster cores, 
  leading to heating of the intracluster medium. 
Dark matter solitons can be formed easily in cosmic walls, 
  and resonance in soliton collision enhances gas trapping and condensation.   
Negative amplitude solitons tend to be larger in size than positive amplitude solitons. 
The locally co-existence of the two types of solitons 
  implies the possibility of filament formation in cosmic voids 


\bigskip
\bigskip
\noindent {\bf DISCUSSION}

\bigskip
\noindent {\bf WOLFGANG KUNDT:} 
Are you familiar with Tsallis Entropy? 
It is a generalisation of the entropy of the ordinary entropy of non-additive systems, 
  which is called ``non-extensive'' by him. 

\bigskip
\noindent {\bf KINWAH WU:} 
Yes, I know Tsallis' work on entropy and thermodynamics. 
We can incorporate part of Tsallis' idea 
  through the assigned degree of freedom in the polytropic equation of state 
  used in our formulation.  

\bigskip
\noindent {\bf JIM BEALL:} 
Can you expand on your perturbative analysis of the system? 
Will the presence of an AGN jet change your calculations? 

\bigskip
\noindent {\bf KINWAH WU:} 
It is possible to expand the perturbative analysis. 
What we need is to consider higher-order terms  
  and keep the relevant equations. 
As for the presence of AGN jets, it will not change the formulation. 
We can treat the AGN jets as an energy source field.   
The energy injected by the jets will change the thermal structures of the galaxy clusters, 
  which is certain.   

\end{document}